# Light-Sail Photonic Design for Fast-Transit Earth Orbital Maneuvering and Interplanetary Flight

Ho-Ting Tung[1], Artur Davoyan[1],*

[1]Mechanical and Aerospace Engineering Department,
University of California Los Angeles, Los Angeles, California, USA.

*E-mail: davoyan@ucla.edu

**Abstract**

Space exploration is of paramount importance to advancing fundamental science and providing global services, such as navigation and communications. However, today's space missions are hindered by limitations of existing propulsion technologies. Here, we examine the use of laser-driven light-sailing for agile Earth orbital maneuvering and for fast-transit exploration of the solar system and interstellar medium. We show that laser propulsion becomes practical at laser powers $\geq 100\ kW$ and laser array sizes ~1 m, which are feasible in the near term. Our analysis indicates that lightweight (1 g – 100 g) wafer-scale (~10 cm) spacecraft may be propelled by lasers to orbits that are beyond the reach of current systems. We further compare our findings with previous interstellar laser propulsion studies, and show that our approach is less constricting on laser architecture and spacecraft photonic design. We discuss material requirements and photonic designs. We show that light-sails made of silicon nitride and boron nitride are particularly well suited for discussed applications. Our architecture may pave the way to ubiquitous Earth orbital networks and fast-transit low-cost missions across the solar system.

**Introduction**

Interplanetary and deep space missions shed light on solar system formation, universe evolution, and, potentially, existence of life [1-3]. Nevertheless, present day space exploration is hindered by fundamental limitations of current propulsion systems [4-6]. Indeed, missions to outer planets may take years of flight time and decades of costly spacecraft development. For example, it took Juno – recent NASA's mission to Jupiter – ~5 years of flight to reach the destination [7]. Planets beyond Saturn have been visited only once [1, 3, 8]. Future space exploration necessitates fundamentally different principles of propulsion and spacecraft design that will significantly reduce flight time, as well as development time and cost.

Spacecraft propulsion limitations are manifested at Earth orbit as well. Massive orbital constellations being deployed to prove global services (e.g., for Earth observations and internet) are driving the economic growth [9]. At the same time, rapid proliferation of orbital use creates novel opportunities and faces new challenges, including on-orbit servicing of satellites on demand [10, 11] and collision avoidance [12, 13], among others. Yet, at present there are no efficient ways to changing spacecraft orbit after launch, such as performing orbital plane or altitude changes maneuvers [4, 5]. In particular, current chemical and electrical rocket engines require either exorbitant amount of fuel or many months of flight time to perform desired orbital maneuvers [4, 5]. A fast and responsive orbital transfer system is yet to be developed.

Here, we study theoretically laser-driven light sailing and show that laser propulsion is of a great promise not only for envisaged future interstellar missions, but also for agile Earth orbital maneuvering and for fast-transit missions to outer planets. Figure 1 illustrates schematically the laser propulsion for near-Earth and solar system exploration. Specifically, we show that with relatively moderate laser beam requirements – 100 kW - 10 MW of power and 1-10 m array aperture – a lightweight ($\leq 100$ g) wafer-scale spacecraft may be propelled to velocities and regimes that are beyond the reach of conventional space propulsion technologies. In particular, we demonstrate that arbitrary Earth orbital maneuvers may be performed in a matter of minutes and show that >10x faster interplanetary flight as compared to today's space missions is possible. We stress conceptual difference of the regime of operation studied here with previous studies of laser driven light sailing focused on interstellar flight [14-20]. We introduce figures of merit for such regime of laser sailing and analyze corresponding material and photonic designs challenges. We discuss several examples of nanostructured thin films that may meet desired operation criteria. Finally, we discuss promise of laser-driven light-sailing for Earth orbital use and for fast and scalable solar system exploration.

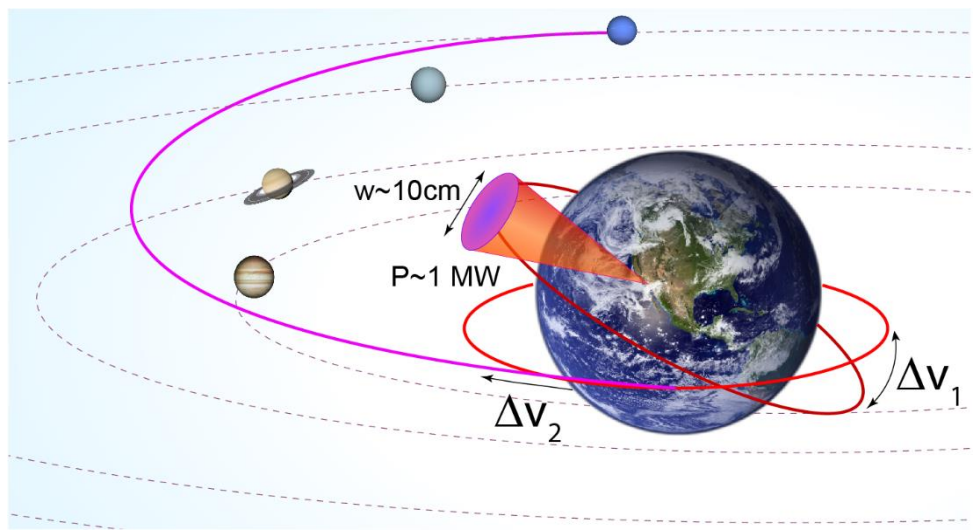

*Figure 1. Schematic illustration of laser sailing for Earth orbital maneuvering and for fast transit interplanetary missions. Here a laser beam with ~1MW power propels a wafer-scale (~10 cm) light-sail. Powered by laser propulsion such a spacecraft can perform highly energetic orbital maneuvers that require very large velocity gain, $\Delta v$. Two examples (not to scale) of such maneuvers are schematically shown: inclination chance at the low Earth orbit and fast transit to Neptune and beyond.*

Laser-driven spacecraft propulsion is an active field of reach in the context of interstellar flight [14-20]. It is expected that with the use of ultrahigh power laser beams pushing large area sails near-relativistic velocities may be reached [16, 18]. For example, a most notable example, is a recently proposed Starshot mission concept [18-20] that poses an audacious goal of sending a probe within a ~20 year timeline to an exoplanet Proxima Centauri b – a potentially habitable planet 4.2 light years away in the closest star system [2]. For this purpose, Starshot proposes to propel a ~1g spacecraft to ~20% of the speed of light by ~100 GW kilometer square laser array [18, 20]. Inspired by the Starshot mission concept we study laser propulsion of light-weight wafer-scale probes in the context of Earth orbital maneuvering and interplanetary flight. As we show below, fundamental advantages of laser propulsion are manifested at much smaller laser powers in applications where conventional electric and chemical rockets are traditionally utilized. Despite an active study of laser propulsion for interstellar travel, we are not aware of works that study in detail laser sailing for Earth orbital and solar system exploration.

**Laser sailing figure of merit**

We begin our analysis by elucidating limitations of conventional electric and chemical rockets and by highlighting unique advantages of laser sailing. Consider an Earth orbiting spacecraft. Its dynamics is governed by the interplay of Earth gravitational attraction $\boldsymbol{F}_g = -\frac{\mu_E m}{r^3}\boldsymbol{r}$ and thrust, i.e., propulsive force, $\boldsymbol{F}_p$ [5] (here $\mu_E$ is a gravitational parameter of Earth, $m$ is the spacecraft mass (note that for regular electrical or chemical rockets $m = m(t)$), and $\boldsymbol{r}$ is the radius vector defining spacecraft position at any given time). In the absence of thrust (i.e., $\boldsymbol{F}_p = 0$) spacecraft follows a familiar Keplerian orbital path [5]. With an applied thrust, $\boldsymbol{F}_p$, orbital changes or space maneuvers, such as, increasing spacecraft altitude, changing orbital plane, or putting a spacecraft onto interplanetary and interstellar trajectories (see Fig. 1) become possible. The ability of a spacecraft to perform orbital maneuvers may then be described by a scalar velocity gain [5]:

$$\Delta v = \int_0^t \frac{|F_p(t)|}{m} dt, \qquad (1)$$

Velocity gain provides a *single unified figure of merit* that allows comparing performance of physically different space systems across arbitrary missions and application domains. Higher velocity gain, $\Delta v$, implies that a broader range of potential orbital transfers to orbits that deviate significantly from the original one may be performed [5, 21]. For instance, complex orbital transfers require $\Delta v > 3\ km/s$. Transferring a spacecraft from a low Earth orbit (LEO) at 300 km altitude to a geostationary orbit (GEO) at ~35,700 km altitude requires $\Delta v \simeq 3.9\ km/s$, putting a spacecraft from LEO onto Earth escape trajectory for interplanetary flight needs $\Delta v \simeq 3.2\ km/s$, whereas performing 90 deg. orbital plane inclination change necessitates $\Delta v \simeq 11\ km/s$, (see also Fig. 2b).

For conventional spacecraft with propulsion originating from fuel expulsion the $\Delta v$ is

given by Tsiolkovsky's rocket equation [5, 21]:

$$\Delta v = v_{\text{ex}} \ln\left(\frac{m_0}{m_f}\right), \qquad (2)$$

where $v_{ex}$ is the fuel exhaust velocity that is defined by an engine $m_0$ and $m_f$ are the spacecraft mass at the beginning and end of maneuvering, respectively. Eq. (2) clearly shows limitations of current engines in performing complex maneuvers that require high $\Delta v$ ($\Delta v \geq 3 km/s$). In particular, a relatively small exhaust velocity of chemical rockets, $v_{ex} \simeq 3 km/s$ [5], according to Eq. (2) requires exorbitant amount of fuel to perform desired high $\Delta v$ orbital maneuvers. For example, achieving $\Delta v = 10 \ km/s$ would require >96% of spacecraft mass allocated to fuel – an insurmountable constrain for space flight. Indeed, the highest in-space (i.e., once the spacecraft is launched onto an orbit) $\Delta v$ achieved to date was by Magellan spacecraft ($\Delta v \simeq 4$ km/s) on its mission to Venus (Fig. 2a) [22, 23]. Electrical engines [4, 24, 25], while efficient in delivering high $v_{ex}$ ($v_{ex} \simeq 30 km/s$ is typical [24]) and thus high $\Delta v$ with affordable fuel consumption (see Eq. (2)), produce very low propulsive force, $|\boldsymbol{F}_p|$<<1N, and therefore result in lengthy – months long – orbital transfers [26, 27], see Eq. (1). For example, Dawn – the first electrically propelled interplanetary mission, which explored Vesta and Ceres [25], attained $\Delta v = 11 \ km/s$ (Fig. 2a). This is the largest $\Delta v$ achieved to date by a rocket. Nevertheless, it took 90 mN Dawn thrusters ~5.5 years to perform the maneuver. Achieving simultaneously fast transfers and high $\Delta v$ are beyond the reach of either chemical or electrical rockets.

Laser sailing [14-20] fundamentally differs from conventional chemical and electric means of propulsion. Laser sailing harnesses photon radiation pressure from an external laser source. Without a need to carry energy source and propellant on board, even minute forces of radiation pressure may accelerate a light-weight spacecraft to near-relativistic velocities [18-20]. To illustrate the capabilities of laser prolusion we analyze the corresponding velocity gain:

$$\Delta v = \int_0^t \frac{|F_p(t)|}{m} dt = \frac{2P}{mc} t, \qquad (3)$$

where $\boldsymbol{F}_p \simeq \frac{2}{c} R cos(\theta) P \boldsymbol{n}$ is the radiation pressure force for a flat specularly reflecting sail [19], $P$ is the laser power, $c$ is the speed of light, $R$ is the reflectivity of the light-sail, $\theta$ is the angle of incidence, and $\boldsymbol{n}$ is a normal to a light-sail, $t$ is the total illumination time. Here $m = m_s + m_p$ is the overall mass of the laser propelled spacecraft, $m_s$ is the mass of the light-sail, and $m_p$ is the mass of the payload, i.e., all other spacecraft systems. In our analysis for the sake of concept demonstration we assume normal incidence ($\theta = 0$), unity reflectivity ($R = 1$) and negligible absorbance at the laser wavelength.

Clearly, a light-weight spacecraft under a high enough laser power can attain very high velocity gain, $\Delta v$, in a relatively short period of time. Figure 2a compares $\Delta v$ possible with $P/m = 1 MW/g$ (compare to $\sim 100 \ GW/g$ considered in Starshot program [20]) with that possible with the state-of-the-art electric and chemical engines. In 600 s (~10 minutes) of illumination, the light-driven spacecraft gains $\Delta v$ surpassing that of best chemical rockets [22, 23]. After 2000 s of illumination $\Delta v$ exceeds that of Dawn attained in 5.5 years [25]. Figure 2b further examines capability of laser-driven light-sailing. Evidently, very high $\Delta v$ needed to perform arbitrarily complex maneuvers may be reached in a relatively short time (minutes to hours) with even moderate laser power requirements $\simeq 100\ kW/g$.

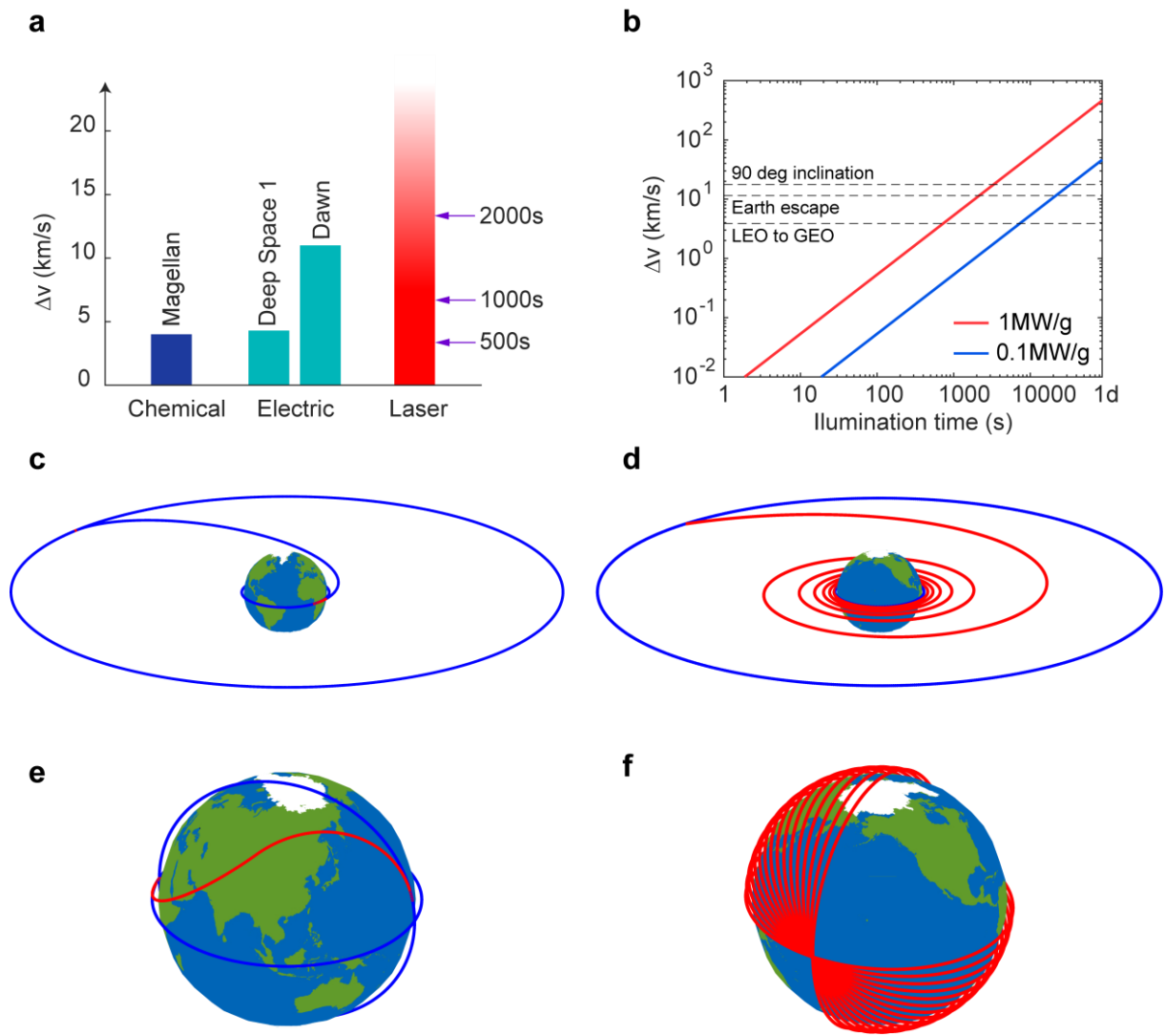

*Figure 1. Comparison of laser sailing to chemical and electric propulsion. **a**, In-space velocity gain attained by electrical and chemical engines, and laser sailing. **b**, Velocity gain with illumination time for two different values of laser power per spacecraft weight. Dashed lines indicate velocity gain to perform respective orbital maneuvers. **c** and **d**, Calculated orbital transfers form LEO to GEO for 1 MW/g and 10 kW/g, respectively. **e** and **f**, Calculated 90deg inclination change maneuvers for 1 MW/g and 34 kW/g, respectively. In **c-f** highlighted parts of the trajectory denote phases of active laser propulsion.*

We continue our analysis with a study of laser sailing use for Earth orbital maneuvering [5]. In particular, we consider two characteristic orbital maneuvers: a transfer from a low Earth orbit to a geostationary orbit and a 90 degree orbital plane inclination change maneuver. Provided that the range of laser operation is limited at most by the geostationary orbit ($z_{GEO} \simeq 35,700\ km$ altitude), we assume that all of the laser power may be focused on the light-sail. Such assumption is justified when laser aperture diameter, $D$, and light-sail radius, $w$, meet Rayleigh length criterion [20, 28]: $Dw > \frac{2}{\pi} z_{GEO} \lambda$, where $\lambda$ is the laser wavelength. For instance, a sail with $w = 1\ m$ would require a laser with an aperture $D \simeq 26\ m$ (compare with the 30m diameter primary mirror of the Thirty Meter Telescope under construction). However, we stress that most practical scenarios are limited to low and medium Earth orbits that require a much shorter operation range ($z \leq 1000 km$), and therefore a significantly smaller laser array. For the sake of concept demonstration, an idealistic scenario of normal incidence and perfect reflectivity (i.e., $\theta = 0$ and $R = 1$) is assumed. Figures 2c and 2e show calculations of altitude and plane change transfers for 1 MW/g, respectively. Fast, quasi-impulsive transfers on par with chemical rockets are possible in this case [5, 21]. It takes total of 583 s of illumination time (~10 min) to insert a spacecraft onto LEO to GEO transfer orbit (Fig. 2c). A 90 deg. plane change maneuver requires even higher $\Delta v \simeq 10\ km/s$ which results in a longer illumination time needed (total of ~40 minutes assuming 1 MW/g). Note that such plane change maneuvers are beyond the reach of chemical rockets at present. With higher power per spacecraft mass (i.e., larger $P/m$ ratio) even faster transfers are possible. On the contrary, by relaxing the transfer time constraint, orbital maneuvers may be performed with lower power lasers (see also Fig. 2b). Hence, by requiring that a transfer is accomplished with total of 1 day of illumination time, we find that a $m = 10\ g$ spacecraft can be transferred from LEO to GEO by $P \simeq 100\ kW$ laser (Fig. 2d), while 90 deg. orbital plane change maneuver (Fig. 2f) is possible with $P \simeq 340\ kW$ laser. Slow transfer trajectories in Figs. 2d and 2f resemble those attained by electric engines [21, 26], however, unlike electric rockets these maneuvers are performed within 1 day (compare with Dawn spacecraft that took >5 years to attain $\Delta v \simeq 11\ km/s$ [25]). Calculations presented in Fig. 2 evidently show that laser propulsion offers a fundamentally different regime of operation, beyond the reach of electrical and chemical rockets. We note that our study pertains to an idealistic scenario. In practice one should consider effects of oblique incidence, Earth rotation and orbital orientation with respect to a laser – effects that depend on a particular mission and its objectives. Study and analysis of these effects is beyond the scope of this manuscript, in which we aim to consider generic aspects of laser sailing for orbital and interplanetary flight. Nonetheless, our analysis provides a guideline for the efficiency of laser propulsion as compared to other means of space propulsion.

**Light-sail photonic design and materials**

Orbital maneuvers discussed in Fig. 2c-2f require use of relatively high power lasers: ~100kW – 1MW to propel 1-10 g spacecraft. Notably this regime of operation is drastically different from that of Starshot [18-20]. Specifically, a more than 4 orders of magnitude smaller laser power puts a different constrain on light-sail design [29-33], suggesting conceptually different performance metrics. Lower laser power implies that a wider class of materials may be used, beyond those considered in Starshot program [19]. Whereas shorter operation distances suggest that smaller area sails with a radius on the order of 10 cm, and, hence, with a lower mass, $m_s$, may be used. In fact, the concept of operation considered here suggests that $m_s \ll m_p$, so that larger fraction of the spacecraft mass is the payload mass $m_p$ (i.e., $\frac{m_p}{m_p+m_s} \to 1$), i.e., mass allocated for instruments and other spacecraft systems. For example, $1\ \mu m$ thick w $w = 10\ cm$ light-sail would weigh less than 100 mg, $m_s \leq 0.1\ g$ (i.e., <10% of total spacecraft mass $m$). To further illustrate this point in the Fig. 3a we plot the velocity gain, $\Delta v$, after $t = 1000s$ of illumination for an idealistic scenario, Eq. (3), as a function of incident laser power and $m_p/m_s$ ratio for a total spacecraft mass of $m_s + m_p = 1g$. For $m_s < 0.1 m_p$ contribution of the light-sail mass to achieving a desired $\Delta v$ strongly diminishes. Small fraction of the light-sail mass provides flexibility in selecting light-sail materials and photonic design. Indeed, a proper figure of merit for light sail design obtained from Eq. (3) is:

$$\begin{cases} \Delta v = \frac{2}{c} Pt \max \frac{R(m_s)}{m_s+m_p}, \\ T < T_c \end{cases} \quad (4)$$

where $T < T_c$ condition ensures that the sail temperature, $T$, is kept below a desired threshold, $T_c$. In the limit of $m_s \ll m_p$ these conditions may be considered independently, implying that the figure of merit simplifies to searching structures that yield simultaneously high reflectivity $R(m_s) \to 1$ to maximize radiation pressure and high emissivity for radiative cooling to desired operation temperature $T < T_c$, while keeping $m_s \ll m_p$. In the following analysis we discuss materials selection and photonic designs that can meet these conditions, namely high reflectivity and low operating temperature in an ultrathin film limit.

In order to identify materials and photonic designs that can meet our metrics, we first study the effects of laser absorption on sail temperature to determine the range of admissible laser absorptivities and thermal emissivities that ensure keeping the light-sail at a low operation temperature. At the equilibrium, sail temperature is found as a power balance between absorbed laser power $P_{abs} = \alpha P$ and power emitted as thermal radiation into the free space $P_{rad} = 2\sigma\epsilon T^4 \pi w^2$:

$$T = \left(\frac{\alpha}{2\epsilon} \times \frac{P}{\pi w^2} \times \frac{1}{\sigma}\right)^{1/4}, \quad (5)$$

where $\alpha$ is the sail absorptivity at the laser wavelength $\lambda$ (nonrelativistic dynamics ensures that Doppler shifting may be neglected in our case), $\epsilon$ is a hemispherical sail emissivity, factor of 2 accounts for front and backside thermal emission (for an ultrathin film sail its front and rear sides within a first order approximation exhibit similar emissivity values [29]), $\pi w^2$ is the sail area, and $\sigma$ is the Stefan–Boltzmann constant. Evidently for a given laser power, $P$, which is chosen to satisfy desired mission requirements (Fig. 2b), the sail temperature depends on $\alpha/\epsilon$ ratio. In Fig. 3b we plot sail temperature variation with the incident laser power, $P$, and light-sail absorptivity to emissivity ratio, $\alpha/\epsilon$, for a sail with $w = 10\ cm$ (i.e., wafer scale [34, 35]). Clearly in the range of powers that are of interest here ($P \leq 1\ MW$) sail temperature may be kept below $T \leq 500\ K$ for $\frac{\alpha}{\epsilon} < 10^{-3}$, which may be reached with a proper choice of materials. Hence, with the use of low loss dielectrics (e.g., $TiO_2$, $Al_2O_3$, $MgF_2$, BN) or wide bandgap semiconductors (e.g., Si, $Si_3N_4$, $MoS_2$, GaAs, C) below their absorption band, absorptivities on the order of $10^{-4}$ to $10^{-6}$ may be attained). In addition, nanophotonic engineering may allow design of thin thermally emissive surfaces with a relatively high emissivity ($\epsilon > 0.1$) [36-41]. Together with $\alpha/\epsilon$ ratio light-sail temperature depends on its area $\pi w^2$: larger area yields higher radiated thermal power and thus lower light-sail temperature for a given incident laser power $P$. Figure 3c plots minimum light-sail radius, $w$, needed to keep its temperature below $T = 500\ K$ as a function of the incident laser power, $P$, for different $\alpha/\epsilon$ ratios. In all of the studied cases the area of a sail needed to radiatively cool it to moderate temperatures is rather small, i.e., $w \propto 10\ cm$. This suggests that small, wafer scale, spacecraft can be launched and propelled by laser beams to perform desired breakthrough fast transit missions.

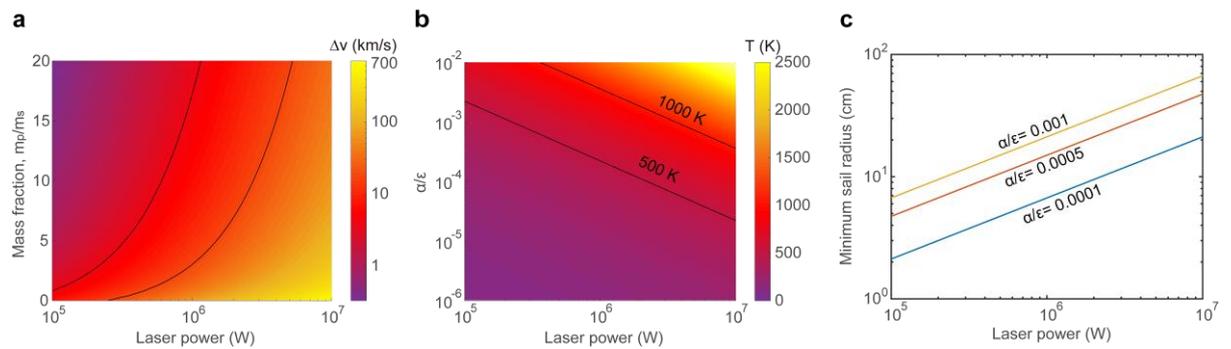

*Figure 3. Light-sail design figures of merit. **a**, Velocity gain, $\Delta v$, with laser power and payload to sail mass ratio. 1000 s of total illumination time is assumed. The curves indicate velocity gain needed to perform LEO to GEO transfer and 90 deg. orbital plane inclination change (see also Fig. 2). **b**, Light-sail temperature with the laser power and $\alpha/\epsilon$ ratio. **c**, Minimum sail area needed to keep the sail at 500 K with laser power variation for different $\alpha/\epsilon$ ratios.*

After determining the range of $\alpha/\epsilon$ needed for a successful operation we proceed with identifying proper materials and photonic designs that can provide $R \simeq 1$ while offering low $\alpha/\epsilon$ ratio and a small light-sail mass. We assume laser wavelength of $\lambda = 1.06\ \mu m$, which is dictated by the atmospheric transparency [36] and availability of low cost high power fiber lasers [42]. Transparent dielectrics, such as $SiO_2$, possess ultra-low loss at the laser wavelength, however, their low refractive index ($n \leq 1.5$) limits options for photonic design resulting in thicker and heavier structures. High refractive index materials ($n \simeq 3.5$), such as Si or GaAs, on the contrary, allow design of a diverse range of ultrathin functional photonic structures [43], nevertheless higher loss at the laser wavelength and weak absorbance in the infrared lead to higher operation temperatures. As on optimal materials for light-sail design we study silicon nitride and boron nitride, which although possessing lower refractive index when compared to Si, ($n \sim 2$ vs $n \simeq 3.5$), are well suited for our needs. Specifically, crystalline stoichiometric silicon nitride ($Si_3N_4$) [44-46] and hexagonal boron nitride (hBN) [47, 48] possess ultralow loss at 1060nm (absorption coefficient $\sim 11\ m^{-1}$). In addition these materials and their allotropies exhibit high infrared absorptivity [44, 49], which offers an efficient pathway for light-sail radiative cooling [39]. In contrast to Starshot LightSail which is constrained by a very stringent mass budget [19], design constrains in the context of the present discussion are significantly relaxed, as soon as $m_s \ll m$ criterion is satisfied. In this case light-sail reflectivity and its thermal emissivity can be optimized independently, as is conceptually illustrated in Fig. 4a.

Figures 4b and 4c show two examples of light-sail reflector designs: Bragg stack and guided mode resonance (GMR) reflectors [50], respectively. We optimize both structures to obtain a near unity reflectivity ($R \simeq 1$) at the target wavelength ($\lambda = 1.06\ \mu m$). Bragg reflector designs (Fig. 4b) have a total thickness $\sim 1.3\ \mu m$ (silicon nitride reflector is comprised of 133 nm thick $Si_3N_4$ layer with 265 nm thick spacers between them; boron nitride design is made of 120 nm thick hBN layers with 265 nm thick spacer layers). Already 4-layer Bragg structures exhibit broad reflection band $0.9\mu m - 1.3\mu m$ with near unity reflectance ($R > 0.98$) at the laser wavelength ($\lambda = 1.06\ \mu m$). GMR structures (Fig. 4c), in turn, enable much thinner, and hence lighter, reflector structures. Specifically, optimal structures with near unity reflectance ($R > 0.99$) at $1.06\ \mu m$ have thickness of about 250 nm. Nonetheless, being resonant in nature [50] GMR designs exhibit much narrower reflectance bandwidth. Low intrinsic optical absorption in $Si_3N_4$ and hBN at the laser wavelength, results in low laser light absorbance. Hence, nonresonant Bragg reflector structures exhibit $\alpha \simeq 10^{-6}$, whereas resonant light-structure interaction in GMR reflectors results in a 100 fold enhanced laser absorption, i.e., $\alpha \simeq 10^{-4}$. We note that many other structures may be used to achieve near-unity reflectivity as well, including metasurfaces [51], photonic crystal membranes [52], and more sophisticated designs obtained by advanced optimization techniques [53]. Our analysis may be generalized to these structures as well.

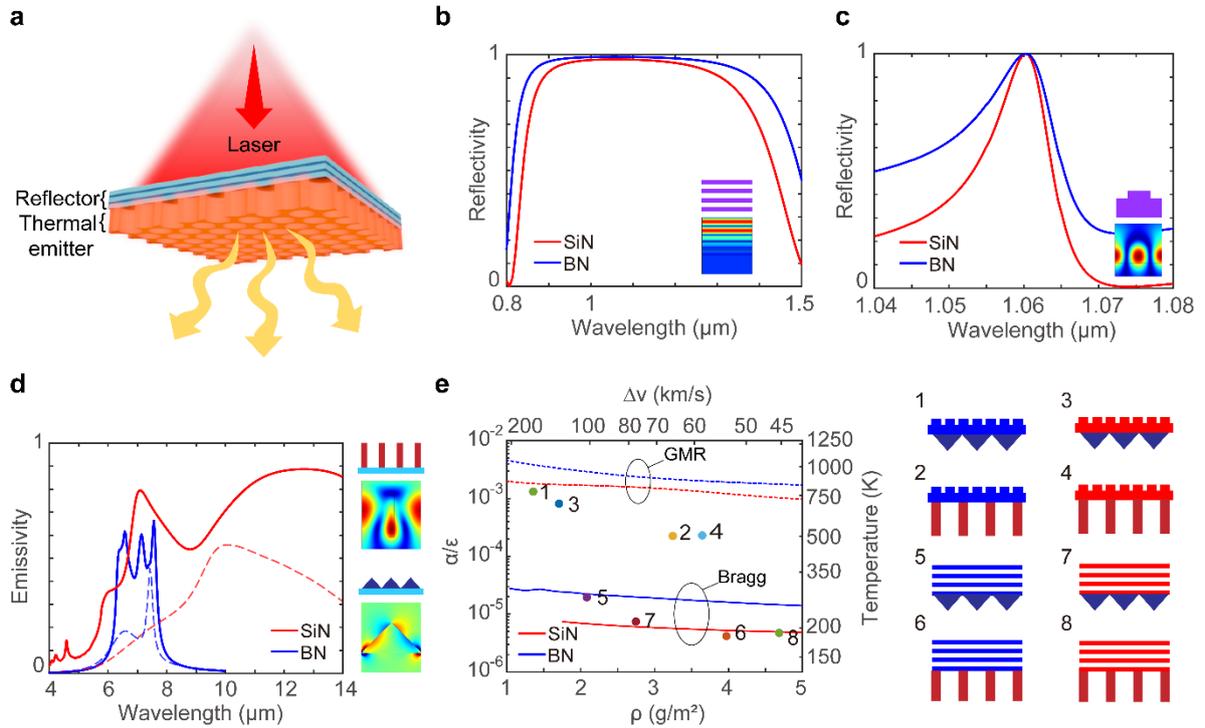

*Figure 2. Light-sail photonic design. **a**, Conceptual illustration of the lightsail depicting an reflector layer that faces the incident laser beam and the thermal emitter layer that dissipates the heat radiatively. **b** and **c**, calculated reflected spectra for $Si_3N_4$ and hBN Bragg reflector and guided mode resonance reflector designs. Insets show respective geometries and calculated electric field intensity profiles at $\lambda = 1.06 \mu m$. **d,** Calculated spectral infrared emissivity spectra for $SiN_x$ and BN thermal emitter designs. Dashed curves denote spectral emissivity of unstructured 500 nm thick BN and $1~\mu m$ thick $SiN_x$ films, respectively. Insets show schematic of the structure and electric field intensity profiles within one unit cell plotted at $7.7~\mu m$ and $6.6~\mu m$ for $SiN_x$ (top) and BN (bottom) emitters, respectively. **e,** Calculated figures of merit for 8 light-sail designs obtained by different permutations of materials and reflector – emitter designs. Insets show schematically respective structures.*

Next we study designs that yield high thermal emissivity. The mechanism of light absorption in silicon nitride and boron nitride is drastically different. Stoichiometric silicon nitride, i.e., $Si_3N_4$, is ultralow loss up-to $\simeq 8.3~\mu m$ and is not suitable for design of efficient thermal emitters [44, 45]. At the same time Si rich silicon nitride, i.e., $SiN_x$, exhibits strong absorbance starting from $\sim 5\mu m$, which makes it well suited for radiative cooling across mid-infrared band [44]. Yet, a careful account of laser absorbance in non-stoichiometric $SiN_x$ should be taken. In our case the thermal emitter layer is nearly completely shielded by the near-unity reflector (Fig. 4a), and hence laser absorbance in the thermal emitter can be neglected. Boron nitride, in turn, exhibits strong polaritonic resonance at $7.3\ \mu m$ (and at $\simeq 13\mu m$), which is associated with formation of a Reststrahlen band with negative material permittivity in a range $6\mu m - 7\mu m$ [47, 48]. Phonon-polariton resonances in hexagonal boron nitride exhibit very high quality factors, which makes coupling of thermal radiation with a crystalline hBN challenging. Broader resonances are observed in boron nitride nanotubes (BNNT), which makes them better suited for thermal emission management [49]. In Fig. 4d we plot spectral emissivity $\epsilon_\lambda$ for thin BNNT (500 nm thick) and Si-rich SiNx (1 $\mu m$ thick) films, respectively. The emissivity can be further enhanced by making use of micro and nanostructured surfaces. Specifically, by perforating the SiNx film (0.265 fill factor) we are able to achieve significant enhancement of emissivity (Fig. 4d) in a broad spectral range, $\lambda > 6\ \mu m$. For BNNT films emissivity enhancement can be attained by resonant excitation of surface phonon-polaritons [47, 48]. In Fig. 4d we plot spectral emissivity of tapered surface phonon-polariton microresonators, which exhibit broad band absorbance across entire Reststrahlen band. Strong and broadband spectral emissivity of both $SiN_x$ and BNNT films implies that these films may be kept at a reasonably low operating temperature even under high power laser irradiation.

We remind that the relaxed constrain on the light-sail mass allows designing reflective and emissive layers independently (Fig. 4a). A final light-sail design is then made of a combination of reflector (e.g., GMR or Bragg) and an emitter layers, respectively. This yields total of 8 possible designs assuming that different material permutations are also accounted. Moving forward, for each of these 8 light-sail designs we study aerial density, $\rho$, and absorptivity to emissivty ratio, $\alpha/\epsilon$. We note that both of these parameters directly translate to the overall figure of merit (Eq. (4)). To be more specific we assume $P = 1\ MW$ of laser power, $w = 10\ cm$ light-sail radius, $m_p = 1g$ payload, and $t = 1000s$ illumination time, and calculate light-sail temperature, $T$, and spacecraft velocity gain, $\Delta v$. The temperature of the sail is found from the radiation balance $P_{rad}(T) = \alpha P_{laser}$, where $P_{rad} = 2\pi w^2 \pi \int_0^\infty \epsilon_\lambda(\lambda) I_{BB}(T,\lambda) d\lambda$, $I_{BB}(T,\lambda) = \frac{2hc^2}{\lambda^5} \frac{1}{e^{hc/\lambda k_B T} - 1}$ is the blackbody radiation at temperature $T$, $k_B$ is the Boltzmann constant, $h$ is the Planck constant, and $c$ is the speed of light, factor $\pi$ approximates hemispherical emission and factor 2 accounts for a double sided emission. The emissivity $\epsilon$ at temperature $T$ is founds as $\epsilon = \frac{\int_0^\infty \epsilon_\lambda(\lambda) I_{BB}(T,\lambda) d\lambda}{\int_0^\infty I_{BB}(T,\lambda) d\lambda}$.

Figure 4e depicts all 8 point designs studied here on a figure of merit plot, i.e. on $(\Delta v, T)$ $(\rho - \alpha/\epsilon)$ plot. GMR reflectors exhibit 2 orders of magnitude higher laser absorptivity, $\alpha$, than Bragg reflectors, which results in a higher light-sail temperature. Thus, light-sails with GMR reflectors demonstrate higher operating temperature (500K - 750K) as compared to Bragg reflectors design ($T < 300K$). At the same time, light-sails with GMR reflectors are lighter (as GMR reflectors are ~4x thinner than respective Bragg reflectors), which, in turn, results in a higher $\Delta v$. Broadband spectral emissivity of silicon nitride (Fig. 4c) results in a better heat rejection (i.e., lower temperature) as compared narrow band BN thermal emitters. However, boron nitride being lighter than silicon nitride allows design of very light-weight light-sails, which eventually translates onto higher velocity gain, $\Delta v$.

**Interplanetary exploration with laser sailing**

As a final illustration of promise of laser sailing we study fast-transit interplanetary and deep space exploration with light-weight laser driven probes. Launching any spacecraft onto an interplanetary mission or solar escape trajectory requires placing a spacecraft onto hyperbolic Earth escape trajectory [5]. Once on a hyperbolic Earth escape trajectory a spacecraft will reach the "edge" of the Earth's gravity well (i.e., boundary of Earth's sphere of influence at $\sim 9.29 \times 10^6\ km$) and enter interplanetary medium where its astrodynamics is dominated by the solar gravity, Fig. 5a. In the context of interplanetary and deeps space travel it is the velocity at the Earth's sphere of influence boundary, $v_{inf}$, that dictates spacecraft's capability for exploring far reaches of space: the higher is the velocity the further out and the faster the spacecraft will reach its target [1, 3, 5, 8, 54]. In Fig. 5b we plot $v_{inf}$ with laser power, P, and laser operation range, $z = \frac{\pi}{2}\frac{Dw}{\lambda}$ [28], that is, a distance over which light-sail can be continuously accelerated. Here we assume $w = 10\ cm$ light-sail and $m = 1g$ overall spacecraft mass. For a $w = 10\ cm$ light-sail the operation ranges of interest correspond to laser apertures with diameters ranging from $D = 2m$ to $D = 50m$. We observe that even with these moderate conditions (as compared to previous studies [19, 20]) velocities exceeding that of solar system escape [3, 5] can be reached (i.e., $v_{inf} > 12.2\ km/s$). Specifically, velocities 5 times higher than that of the New Horizon spacecraft [8], a recently launched probe for Pluto and Kuiper exploration, are possible. Clearly, with such high velocities new opportunities for solar system exploration emerge. Important to note that the actual transit time to a given planet depends on its phasing with Earth (i.e., mutual positions of planets about the sun at any given moment of time [5, 21]), and can be found by solving a respective astrodynamics problem [21]. Figure 5c depicts calculated transit times for all possible orientations between Earth and other planets of the solar system. With $v_{inf} = 50 km/s$ Mars may be reached in 20 days (cf. with 200 days for Perseverance to arrive Mars), Jupiter in 120 day (cf. with 5 years of the Juno mission [7]), Pluto in 1000 days (cf. with 10 years of the New Horizons [8]), and 100 AU in 10 years (cf. with 45 years of Voyager 1 [3]). The capability offered by laser sailing even at moderate power and spacecraft requirements discussed here may open new avenues for missions to interstellar space and capturing interstellar objects [55]. Such missions may set a ground for future interstellar flight [18, 20].

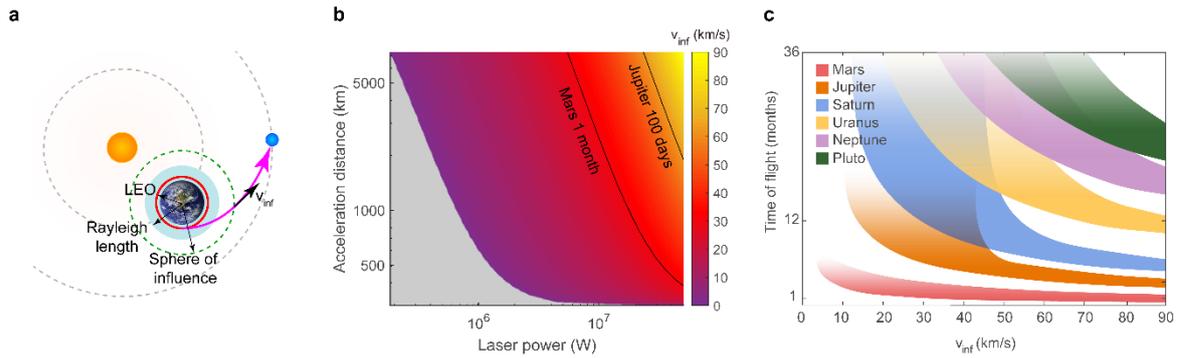

*Figure 3. Light-sailing for fast-transit interplanetary exploration. **a,*** *Schematic illustration (not to scale) of an interplanetary maneuver. Starting from LEO a light-sail is inserted onto a hyperbolic Earth-escape trajectory. **b**, Hyperbolic excess velocity, $v_{inf}$, with laser power and acceleration distance. **c,** Calculated time of flight to different planets of the solar system depending on the value of $v_{inf}$. Here the shaded "bands" correspond to possible planetary orientations with respect to Earth.*

**Conclusions**

In summary, we have discussed a conceptually different regime of operation for laser driven light-sailing. We showed that even with a moderate parameter range laser-driven light sailing might outperform conventional propulsion methods offering novel opportunities for Earth orbital maneuvering and space exploration. We have introduced a figure of merit – $\Delta v$ and have discussed materials and design strategies that can satisfy this figure of merit. We envisage that laser-driven light sailing can pave the way to a new class of space missions for fast space exploration.

**Methods**

Optical properties of structures discussed here were calculated with Lumerical$^{TM}$. The refractive index data for $Si_3N_4$ and hBN at the laser wavelength is adopted from [46] and [56], respectively. To account for the small but non-zero absorption at the laser wavelength, the extinction coefficient for both $Si_3N_4$ and hBN is set to be $10^{-6}$ [57]. The near infrared optical data of $SiN_x$ and BN are from [44] and [49], respectively. In our numerical calculations we assume 2D structures and TE polarized waves. The density of the structures is estimated by extrapolating 2D designs to a 3D structure, which provides a more accurate density estimate.

The interplanetary transit times are found by solving respective Lambert's problem [21, 58].

**References**:
[1] A. A. Siddiqi and R. Launius, (2002).
[2] G. Anglada-Escudé *et al.*, Nature **536**, 437 (2016).
[3] E. Stone, Nature Astronomy **1**, 896 (2017).
[4] D. Krejci and P. Lozano, Proceedings of the IEEE **106**, 362 (2018).

[5] H. D. Curtis, *Orbital mechanics for engineering students* (Butterworth-Heinemann, 2019).

[6] R. J. Litchford and J. A. Sheehy, in *43rd Annual AAS Guidance, Navigation and Control Conference* (Breckenridge, CO, 2020).

[7] S. Bolton *et al.*, Space Science Reviews **213**, 5 (2017).

[8] A. Witze, Nature News **523**, 140 (2015).

[9] G. Curzi, D. Modenini, and P. Tortora, Aerospace **7**, 133 (2020).

[10] A. M. Long, M. G. Richards, and D. E. Hastings, Journal of Spacecraft and Rockets **44**, 964 (2007).

[11] M. Luu and D. E. Hastings, in *ASCEND 2020* (2020), p. 4127.

[12] N. Sánchez-Ortiz, M. Belló-Mora, and H. Klinkrad, Advances in Space Research **38**, 2107 (2006).

[13] C. Bonnal, J.-M. Ruault, and M.-C. Desjean, Acta Astronautica **85**, 51 (2013).

[14] G. Marx, Nature **211**, 22 (1966).

[15] J. Redding, Nature **213**, 588 (1967).

[16] R. L. Forward, Journal of Spacecraft and Rockets **21**, 187 (1984).

[17] G. A. Landis, (1989).

[18] P. Lubin, arXiv preprint arXiv:1604.01356 (2016).

[19] H. A. Atwater *et al.*, Nature materials **17**, 861 (2018).

[20] K. L. Parkin, Acta Astronautica **152**, 370 (2018).

[21] D. A. Vallado, *Fundamentals of astrodynamics and applications* (Springer Science & Business Media, 2001), Vol. 12.

[22] R. Saunders *et al.*, Journal of Geophysical Research: Planets **97**, 13067 (1992).

[23] K. Hibbard, L. Glaze, and J. Prince, Acta Astronautica **73**, 137 (2012).

[24] M. D. Rayman and D. H. Lehman, Acta Astronautica **41**, 289 (1997).

[25] J. Brophy *et al.*, in *39th AIAA/ASME/SAE/ASEE Joint Propulsion Conference and Exhibit* 2003), p. 4542.

[26] H. R. Kaufman and R. S. Robinson, Journal of Spacecraft and Rockets **21**, 180 (1984).

[27] B. M. Kiforenko, Z. V. Pasechnik, and I. Y. Vasil'ev, Acta Astronautica **60**, 801 (2007).

[28] P. E. Nielsen, Effects of directed energy weapons, (1994).

[29] O. Ilic, C. M. Went, and H. A. Atwater, Nano letters **18**, 5583 (2018).

[30] W. Jin, W. Li, M. Orenstein, and S. Fan, ACS Photonics **7**, 2350 (2020).

[31] M. M. Salary and H. Mosallaei, Laser & Photonics Reviews **14**, 1900311 (2020).

[32] J. Brewer *et al.*, arXiv preprint arXiv:2106.03558 (2021).

[33] M. M. Salary and H. Mosallaei, Advanced Theory and Simulations **4**, 2100047 (2021).

[34] D. J. Barnhart, T. Vladimirova, A. M. Baker, and M. N. Sweeting, Acta Astronautica **64**, 1123 (2009).

[35] Z. Manchester, M. Peck, and A. Filo, (2013).

[36] A. P. Raman *et al.*, Nature **515**, 540 (2014).

[37] W. Li and S. Fan, Opt. Express **26**, 15995 (2018).

[38] K. Sun *et al.*, ACS photonics **5**, 2280 (2018).

[39] K. Sun *et al.*, Acs Photonics **5**, 495 (2018).

[40] D. G. Baranov *et al.*, Nature materials **18**, 920 (2019).

[41] A. Howes, J. R. Nolen, J. D. Caldwell, and J. Valentine, Advanced Optical Materials **8**, 1901470 (2020).

[42] C. Jauregui, J. Limpert, and A. Tünnermann, Nature photonics **7**, 861 (2013).

[43] A. Arbabi, Y. Horie, M. Bagheri, and A. Faraon, Nature nanotechnology **10**, 937 (2015).

[44] J. Kischkat *et al.*, Applied optics **51**, 6789 (2012).

[45] P. Tai Lin, V. Singh, L. Kimerling, and A. Murthy Agarwal, Applied physics letters **102**, 251121 (2013).

[46] K. Luke *et al.*, Optics letters **40**, 4823 (2015).

[47] J. D. Caldwell *et al.*, Nature communications **5**, 1 (2014).

[48] Z. Jacob, Nature materials **13**, 1081 (2014).

[49] X. G. Xu *et al.*, ACS nano **8**, 11305 (2014).

[50] S. Wang and R. Magnusson, Applied optics **32**, 2606 (1993).

[51] P. Moitra *et al.*, Acs Photonics **2**, 692 (2015).

[52] J. P. Moura *et al.*, Opt. Express **26**, 1895 (2018).

[53] S. Molesky *et al.*, Nature Photonics **12**, 659 (2018).

[54] J. H. Mauldin, NASA STI/Recon Technical Report A **93**, 25710 (1992).

[55] D. Garber *et al.*, arXiv preprint arXiv:2106.14319 (2021).

[56] J. Wu *et al.*, Journal of Applied Physics **119**, 203107 (2016).

[57] R. Kou *et al.*, Journal of Applied Physics **126**, 133101 (2019).

[58] E. Lancaster and R. Blanchard, *A unified form of Lambert's theorem* (National Aeronautics and Space Administration, 1969), Vol. 5368.